\documentstyle[12pt,epsf,epsfig]{article}
\def\be{\begin{equation}}
\def\ee{\end{equation}}
\def\ba{\begin{eqnarray}}
\def\ea{\end{eqnarray}}

\def\bq{\begin{quote}}
\def\eq{\end{quote}}

 at 10truept

\newcommand{\beq}{\begin{equation}}
\newcommand{\eeq}{\end{equation}}
\newcommand{\beqa}{\begin{eqnarray}}
\newcommand{\eeqa}{\end{eqnarray}}

\def\ltap{\ \raise.3ex\hbox{$<$\kern-.75em\lower1ex\hbox{$\sim$}}\ }
\def\gtap{\ \raise.3ex\hbox{$>$\kern-.75em\lower1ex\hbox{$\sim$}}\ }
\def\gl{\ \raise.5ex\hbox{$>$}\kern-.8em\lower.5ex\hbox{$<$}\ }
\def\roughly#1{\raise.3ex\hbox{$#1$\kern-.75em\lower1ex\hbox{$\sim$}}}

\begin{document}

\begin{titlepage}
\vfill
\begin{flushright}
%\today
\end{flushright}

\vfill
%\vskip 1.0cm
\begin{center}
\baselineskip=16pt
{\Large\bf  C-Functions in Lovelock Gravity}
\vskip 1.0cm
{\large {\sl }}
\vskip 10.mm
{\bf Mohamed M. Anber\footnote{\tt
manber@physics.umass.edu} and
David Kastor\footnote{\tt kastor@physics.umass.edu}} \\
%\\[2mm]
\vskip 1cm
%\vfill
{

       Department of Physics\\
       University of Massachusetts\\
       Amherst, MA 01003\\
}
\vspace{6pt}
\end{center}
\vskip 0.5in
\par
\begin{center}
{\bf ABSTRACT}
\end{center}
We present C-functions for static and spherically symmetric spacetimes in Lovelock gravity theories. These functions are monotonically increasing functions of the outward radial coordinate and acquire their minima when evaluated on the horizon. Unlike the case of Einstein gravity, where there is a single C-function, we find that this function is non-unique in the case of Lovelock gravity. We define two C-functions, which agree at the horizon giving the black hole entropy, and state the different energy conditions that must hold in order for these functions to satisfy the monotonicity condition.

\begin{quote}

\vfill
% \hrule width 5.cm
\vskip 2.mm
\end{quote}
\end{titlepage}
%%%%%%%%%%%%%%%%%%%%%%%%%%%%%%%%%%%%%%%%

%
\section{Introduction}

Since the remarkable discovery \cite{Bekenstein:1973ur} that a black hole has entropy proportional to the area of the horizon
\begin{equation}
S=\frac{A_{\mbox{\scriptsize H}}}{4G_4}\,,
\end{equation}
many approaches have been proposed to count the number of quantum mechanical states that contribute to this entropy. Particularly intriguing are connections with two dimensional conformal field theory. In this context Solodukhin \cite{Solodukhin:1998tc} and Carlip \cite{Carlip:1999db} showed that if we consider the black hole horizon as a boundary condition on the radial fluctuations of the metric then we obtain, in the vicinity of the horizon, an infinite-dimensional group of conformal transformations in two dimensions with corresponding Virasoro algebra that contains the Bekenstein-Hawking entropy as a central charge. For quantum theories in two dimensions, Zamolodchikov \cite{Zamolodchikov:1986gt} was able to prove a set of properties satisfied by what is called the C-function under renormalization group flow. This function was shown to be a function of the couplings of the theory that is monotonically decreasing as one flows to lower energies. For fixed points of the flow, corresponding to the extrema of this function, the C-function reduces to the central charge of the Virasoro algebra. In \cite{Alvarez:1998wr} a holographic version of Zamolodchikov's C-theorem was proved by studying the renormalization group flow along null geodesic congruences in asymptotically AdS spaces. Further, Sahakian \cite{Sahakian:1999bd}  proposed a covariant geometrical expression for the C-function for theories which admit a dual gravitational description. In this description, the IR region is deep interior and the flow outward in radius is toward the UV region in the QFT sense. Another possible interpretation of the holographic picture using the moduli flow, in the context of the attractor mechanism, was given in \cite{Astefanesei:2007vh}.

It was shown by Goldstein et al \cite{Goldstein:2005rr} that in 4 dimensional Einstein gravity,  coupled to matter fields that satisfy the null energy condition, one can define a simple C-function for static asymptotically flat solutions. This function is given by
\begin{equation}
C(r)=\frac{A(r)}{4\,G_{4}}\,,
\end{equation}
 where $A(r)$ is the area of the two sphere as a function of the radial coordinate and $G_4$ is Newton's constant in 4-D. It was proved in  \cite{Goldstein:2005rr} that the equations of motion imply that $A(r)$ must decrease as one moves inwards from asymptotic infinity. Also, $C(r)$ coincides with the entropy at the horizon.

\smallskip

On the other hand the entropy in higher curvature gravity is given by the integral of a particular local quantity on a spatial cross section $\Sigma$ of the event horizon \cite{Wald:1993nt, Iyer:1994ys}
\begin{equation}\label{ Wald's formula}
S=-2\,\pi\int_{\Sigma}\frac{\partial L}{\partial R_{abcd}}\epsilon_{ab}\epsilon_{cd}\sqrt{-h}\,d\Omega\,,
\end{equation}
where $L$ is the Lagrangian, $\epsilon_{ab}$ denotes the binormal to the horizon cross section and $\sqrt{h}\,d\Omega$ is the volume element induced on $\Sigma$.

The question of whether one can define analogous C-functions in higher curvature gravity was raised in \cite{Cremades:2006ke}, where it was shown that a similar C-function can be obtained by evaluating Wald's expression for the entropy (\ref{ Wald's formula}) on a general spacelike surface instead of  a spatial cross-section of the event horizon. Although the authors in \cite{Cremades:2006ke} pointed out the monotonicity of the C-function for $f(R)$ gravity (this had been established earlier in  \cite{Jacobson:1995uq}), they were not able to draw a conclusion about the monotonicity of such functions in a general theory of gravity. 

An important class of higher curvature gravity theories is known as Lovelock gravity \cite{Lovelock:1971yv}. These are the most general second order gravity theories in higher dimensional spacetimes. A general formula for the entropy of stationary black holes in these theories was obtained by Jacobson and Myers \cite{Jacobson:1993xs}. It includes a sum of intrinsic curvature invariants integrated over a cross section of the horizon. This entropy coincides with the result one obtains using the Wald's formula 
\footnote{ The black hole entropy for Lovelock AdS gravity can also be obtained directly from a background-independent regularization of the Euclidean action as was shown in \cite{Kofinas:2006hr, Kofinas:2007ns}.}
(\ref{ Wald's formula}).
 In addition, similar calculations to \cite{Solodukhin:1998tc} were performed for Lovelock gravity \cite{Cvitan:2002cs} and it was shown, as in  \cite{Solodukhin:1998tc} for Einstein gravity, that the central charge of the Virasoro algebra is proportional to Jacobson-Myers entropy.

One can ask whether a  C-function similar to that of \cite{Goldstein:2005rr} exists for the static spherically symmetric black holes in Lovelock gravity. In other words, one asks if a monotonically increasing function of the outward radial coordinate may exist under certain conditions, which reduces to the entropy when evaluated on the event horizon.  We address this question in the present work. We show not only that such a C-function exists, but also that this function is non-unique. In fact we find two different C-functions that we call C-functions of the first and second kind.These functions  exist provided that the matter content satisfies respectively the null, as in \cite{Goldstein:2005rr}, and the weak energy condition, and that the spacetime is asymptotically flat. 

In the next section we review the construction of pure Lovelock gravity. Then we review an argument proving the monotonicity of  $A(r)$, the area of concentric spheres, in the spherically asymptotically flat spacetimes. In section 3 we introduce the C-functions of the first and second kind of pure Lovelock gravity theories and we prove the monotonic behavior of these functions. Then in section 4 we consider the case of general Lovelock gravity and the behavior of the general C-functions in this theory. The proof of the monotonicity for C-functions of the second kind in general Lovelock gravity is cumbersome and requires thorough analysis for general polynomials of arbitrary degree. We work out the proof for Gauss-Bonnet gravity and we present numerical results for the second and third order Lovelock theories. These numerical results confirm our analytical results in the Gauss-Bonnet case and indicate that the C-function of the second kind may be monotonic in a general third order Lovelock gravity theory as well.

\section{Lovelock gravity}

The Lagrangian density for general Lovelock gravity in $D$ dimesnions is  ${\cal L}=\sum_{m=0}^{[D/2]}\alpha_{m}\,{\cal L}_{m}$, where  ${\cal L}_{m}$ is given by 
\begin{equation}\label{LUVLagrangian}
{\cal L}_{m}=\frac{1}{2^m}\sqrt{-g}\,\delta_{c_1d_1...c_md_m}^{a_1b_1...a_mb_m}\,R_{a_1b_1}{}^{c_1d_1}\,....\,R_{a_mb_m}{}^{c_md_m}\,,
\end{equation} 
$\alpha_{m}$ is the $m$'th order coupling constant, $[D/2]$ denotes the integer value of $D/2$ and the latin indices $a$,$b$,$c$ and $d$ go from $0$ to $D-1$. The $\delta$ symbol is a totally antisymmetric product of $2m$ Kronecker deltas normalized to take the values of $\pm1$. The term ${\cal L}_{0}=\sqrt{-g}$ is the cosmological term, while ${\cal L}_{1}=\sqrt{-g}\,\delta_{c_1d_1}^{a_1b_1}\,R_{a_1b_1}{}^{c_1d_1}/2$ is the Einstein term. In general ${\cal L}_m$ is the Euler class of a $2m$ dimensional manifold.  

\smallskip
As a special class of general Lovelock gravity, we take a theory with highest order interaction ${\cal L}_m$, $m \le [(D-1)/2]$, and send the coefficients of all the lower order terms to zero. We call these pure Lovelock gravity theories, with pure Einstein gravity as the first non-trivial example \cite{Kastor:2006vw}. There has been intensive effort to study black holes as well as their thermodynamic properties in the context of Lovelock gravity ( see e.g. \cite{Boulware:1985wk}-\cite{Cai:2006pq}).

\smallskip
In the following we will be interested in static spherically symmetric spacetimes. Hence, the metric can be assumed to take the form
\begin{equation}
\label{spherical metric}
ds^{2}=-a(r)^2\,dt^2+\frac{dr^2}{a(r)^2}+b(r)^2\,d\Omega_{n}^2\,,
\end{equation}
where $d\Omega_{n}^2$ is the metric on the unit $n=D-2$ sphere. The nonzero components of the Riemann tensor for the above metric are given by 
\begin{eqnarray}\label{first Reiman}
R_{rt}{}^{rt}&=&-(a^{\prime\prime}\,a+a^{\prime\;2}),\quad
R_{ri}{}^{rj}= - a^2\,({b^{\prime\prime}\over b} + {a^\prime\, b^\prime\over {a\,b}})\,\delta_i^j,\\
R_{ti}{}^{tj}&=& -\frac{a\,a^{\prime}\,b^{\prime}}{b}\,\delta_i^j,\quad
R_{ij}{}^{kl}=\frac{1-a^2\,b^{\prime\; 2}}{b^2}\,\delta_{ij}^{kl}\label{second Reiman}\,.
\end{eqnarray}

The equations of motion following from the Lovelock Lagrangian (\ref{LUVLagrangian}) are given by $\alpha_{m}\,{\cal G}_{e}{}^{f\,(m)}=-2^{m+1}\,T_{e}^{f}$ where
\begin{equation}\label{LUVequations of motion}
{\cal G}_{e}{}^{f\,(m)}=\delta_{ec_1d_1...c_md_m}^{fa_1b_1...a_mb_m}\,R_{a_1b_1}{}^{c_1d_1}\,....\,R_{a_mb_m}{}^{c_md_m}\,,
\end{equation}
and $T_{e}^{f}$ is the energy-momentum tensor. Using eqs. (\ref{first Reiman}) and (\ref{second Reiman}), and through repeated application of the identity
\begin{equation}\label{recurance}
\delta^{a_1...a_p}_{b_1...b_p}\,\delta^{b_{p-1}b_p}_{a_{p-1}a_p}=2(D-p+1)(D-p+2)\,\delta^{a_1...a_{p-2}}_{b_1...b_{p-2}}\,,
\end{equation}
we obtain
\begin{eqnarray}
\nonumber
{\cal G}_t{}^{t\,(m)}&=&-\frac{2^{m+1}m\,n!}{(n-2m+1)!}\,a^2\,\left(\frac{b^{\prime\prime}}{b}+\frac{a^\prime\,b^{\prime}}{a\,b}\right)\,\left(\frac{1-a^2\,b^{\prime\;2}}{b^2}\right)^{m-1}+\frac{2^m\,n!}{(n-2m)!}\left(\frac{1-a^2\,b^{\prime\;2}}{b^2}\right)^m\,,\\
{\cal G}_r^{r\,(m)}&=&-\frac{2^{m+1}m\,n!}{(n-2m+1)!}\,\left(\frac{a\,a^\prime\,b^\prime}{b}\right)\,\left(\frac{1-a^2\,b^{\prime\;2}}{b^2}\right)^{m-1}+\frac{2^m\,n!}{(n-2m)!}\left(\frac{1-a^2\,b^{\prime\;2}}{b^2}\right)^m\,.\label{tt and rr equations of motion}
\end{eqnarray}

\subsection{Monotonicity of $b$ and C-function in Einstein gravity}
As a warm up, we recall how the results of Goldstein et al \cite{Goldstein:2005rr} come about. For $m=1$ we obtain  the special case of Einstein gravity in $D$ dimensions for which eqs. (\ref{tt and rr equations of motion}) take the simple form
\begin{eqnarray}
\nonumber
{\cal G}_{t}{}^{t\,(1)}&=&-4\,n\,a^2\,\left(\frac{b^{\prime\prime}}{b}+\frac{a^\prime\,b^{\prime}}{a\,b}\right)+2\,n\,(n-1)\left(\frac{1-a^2\,b^{\prime\;2}}{b^2}\right)\,,\\
{\cal G}_{r}{}^{r\,(1)}&=&-4\,n\,\left(\frac{a\,a^\prime\,b^\prime}{b}\right)+2\,n\,(n-1)\left(\frac{1-a^2\,b^{\prime\;2}}{b^2}\right)\,.\label{special Einstein gravity}
\end{eqnarray}
Now, consider the coupling of this theory to matter that satisfies the null energy condition, i.e. the energy-momentum tensor satisfies the condition
\begin{equation}
T_{ab}\,\xi^{a}\xi^{b} \ge 0\,
\label{the null energy condition for the energy-momentum tensor}
\end{equation} 
for all null vectors $\xi^{a}$. As a special case, one may take a perfect fluid with energy-momentum tensor given by
\begin{equation}
T_{ab}=(\rho+p)U_{a}U_{b}+p\,g_{ab}\,,
\label{perfect fluid energy-momentum tensor}
\end{equation}
where $\rho$, $p$ and $U_{a}$ are respectively the fluid energy density, pressure and $D$-velocity. In this case the condition (\ref{the null energy condition for the energy-momentum tensor}) reads $\rho+p \geq 0$.
 Goldstein et al \cite{Goldstein:2005rr} showed that in this system $b$ is a monotonically increasing function of $r$ for any static, spherically symmetric and asymptotically flat spacetime. To show this one takes a particular linear combination of  ${\cal G}_{t}{}^{t\,(1)}$ and ${\cal G}_{r}{}^{r\,(1)}$
\begin{equation}
{\cal G}_{t}{}^{t\,(1)}-{\cal G}_{r}{}^{r\,(1)}=-64\,\pi\,G_D\,\left(T_{t}{}^{t}-T_{r}^{r}\right)=64\,\pi\,G_D\,T_{ab}\,\xi^{a}\xi^{b}\ge 0\,,
\end{equation} 
where $\xi^{a}=\left(\xi^t,\xi^r \right)$ are components of a null vector, satisfying the relation, $\left(\xi^{t}\right)^2=-g^{tt}$ and $\left(\xi^{r}\right)^2=g^{rr}$, and we have used $\alpha_1=1/16\,\pi\,G_D$ and $G_D$ is Newton's constant in $D$ dimensions. Using eqs. (\ref{special Einstein gravity}) one obtains
\begin{equation}\label{equation of motion of b in Einstein gravity}
a^2\,b^{\prime\prime}=-\frac{16\,\pi\,G_{D}}{n}\,b\,T_{ab}\,\xi^{a}\xi^{b}\,.
\end{equation}
As long as we are outside the horizon, $a^2>0$, we obtain $b''<0$. Given that the spacetime is asymptotically flat, Goldstein et al \cite{Goldstein:2005rr} were then able to show that $b(r)$ is a monotonically increasing function of $r$. The additional steps required are given below in the context of pure Lovelock gravity.
\smallskip
 
 Using this fact, we see that a possible C-function for Einstein gravity takes the simple form 
 \begin{equation}
 C^{\mbox{\scriptsize E}}=\Omega_{n}\,b^n/4\,G_D=A_{S^n}/4\,G_D\,,
 \end{equation} 
where $\Omega_{n}$ is the volume of the unit $n$ sphere. This expression coincides with the entropy when evaluated on the horizon.

\subsection{Monotonicity of $b$ in pure Lovelock gravity}

\subsubsection*{The null energy condition}

Consider pure Lovelock gravity in  $D$ dimensions and of order $m \leq [(D-1)/2]$ coupled to matter that satisfies the null energy condition. We show that in such a system $b$ is a monotonically increasing function of $r$ for any static, spherically symmetric and asymptotically flat spacetime given the positivity of the coupling constant, i.e. $\alpha_{m}\ge 0$. As we did in the case of Einstein's gravity, we take linear combinations of ${\cal G}_{t}{}^{t\,(m)}$ and ${\cal G}_{r}{}^{r\,(m)}$
\begin{equation}
{\cal G}_{t}{}^{t\,(m)}-{\cal G}_{r}{}^{r\,(m)}=-\frac{2^{m+1}}{ \alpha_{m}}\left(T_{t}{}^{t}-T_{r}^{r}\right)=\frac{2^{m+1}}{ \alpha_{m}}T_{ab}\,\xi^{a}\xi^{b}\ge 0\,.
\end{equation} 
 Using eqs. (\ref{tt and rr equations of motion}) we obtain
\begin{equation}\label{equation of motion of b}
b^{\prime\prime}=-\frac{(n-2m+1)!}{2^{m+1}\,m\,n!}\frac{b}{a^{2}}\left(\frac{b^2}{1-a^2\,b^{\prime\,2}}\right)^{m-1}\left(\frac{2^{m+1 }}{ \alpha_{m}}T_{ab}\,\xi^{a}\xi^{b}\right)\,.
\end{equation}

\begin{figure}[ht]
\leftline{
\includegraphics[width=0.3\textwidth]{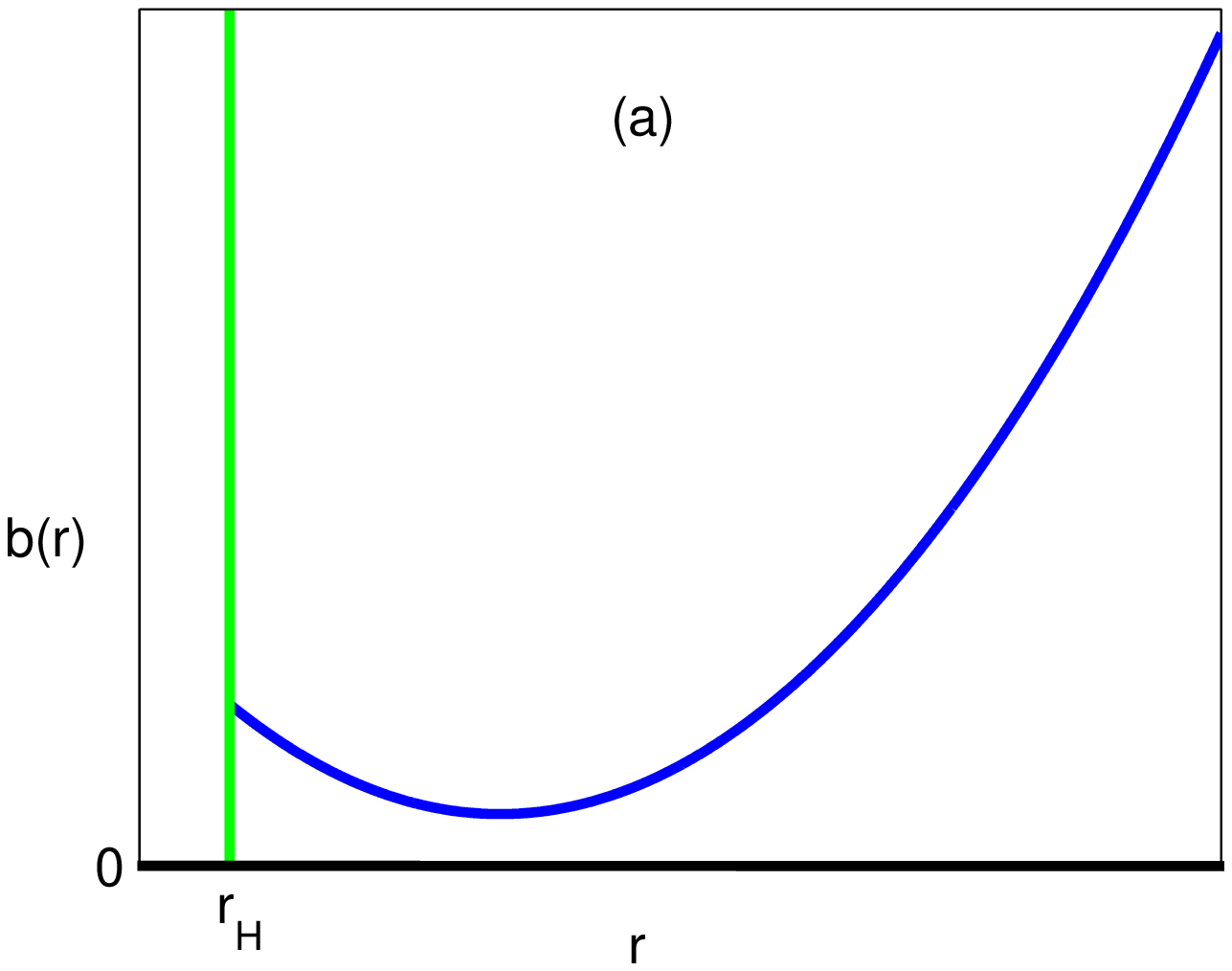}
\includegraphics[width=0.3\textwidth]{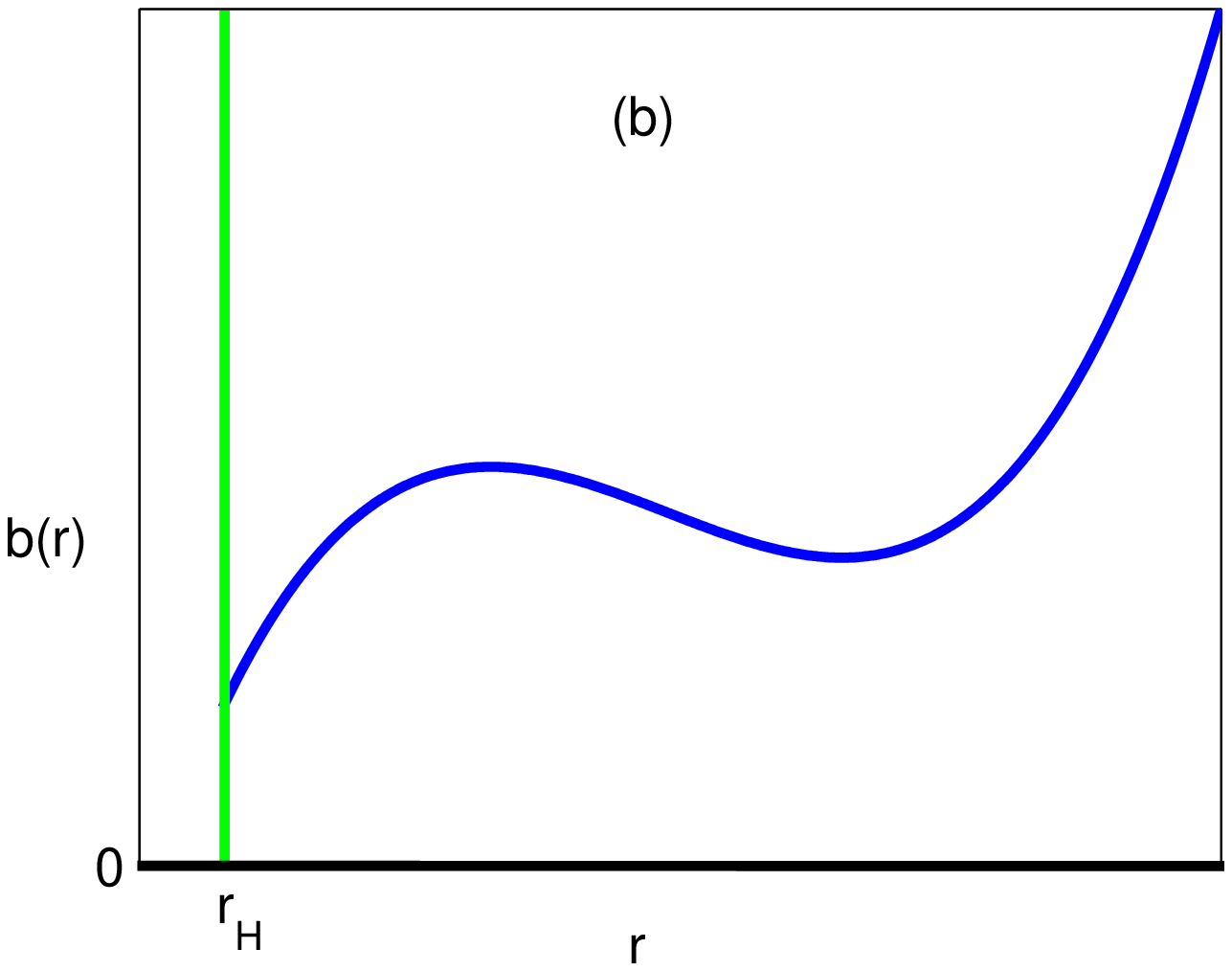}
\includegraphics[width=0.3\textwidth]{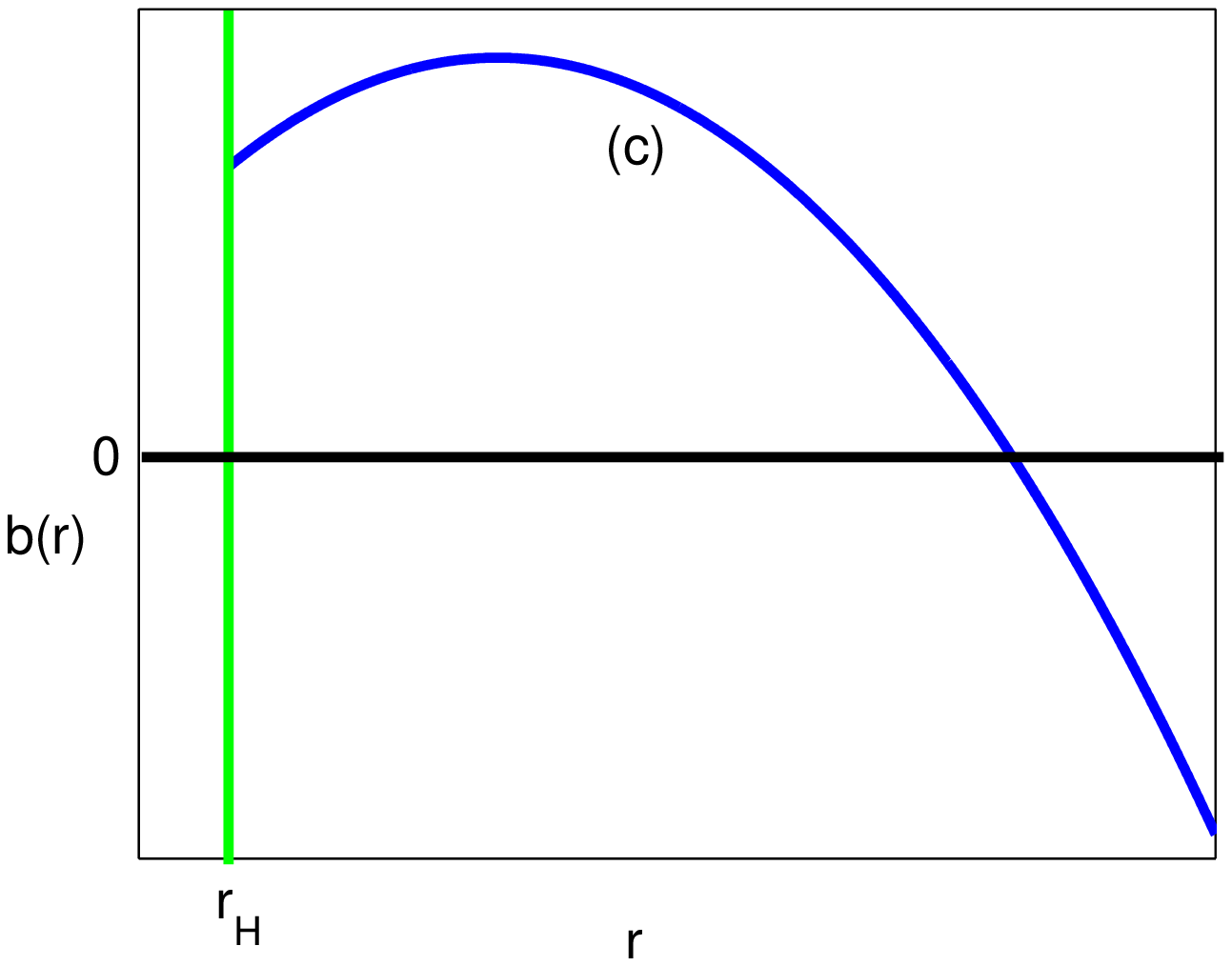}
}

\caption{
Different possibilities for the function $b(r)$.   
}
\label{different possibilities}
\end{figure}

Now, let us assume that the metric (\ref{spherical metric}) describes a black hole. We assume cosmic censorship, so that $b(r)\ne 0$ on, or outside the horizon. Without loss of generality, we can then assume that $b>0$ on the horizon. Asymptotic flatness is consistent with $b(r) \approx \pm r$ as $r \rightarrow \infty$. However, the case $b(r)\approx -r$ depicted on the right  in figure (\ref{different possibilities}) is ruled out by our assumption of cosmic censorship. The other two graphs in figure (\ref{different possibilities}) make clear that for $b(r)$ to then fail to be monotonic between the horizon and infinity, it must have at least one minimum in this range. However, this is ruled out by eq. (\ref{equation of motion of b}). Assume $b^{\prime}=0$ at some radius $r_0$, since $a^2>0$ outside the horizon, it follows from eq. (\ref{equation of motion of b}) that $b^{\prime\prime}(r_0)<0$. Therefore, local minima are not allowed. This proves the monotonicity of $b$ for pure Lovelock gravity.

\subsubsection*{The weak energy condition}
The weak energy condition states that the energy density of any matter distribution, as measured by any observer in spacetime, must be nonnegative, i.e.
\begin{equation}\label{weak energy condition}
T_{ab}\chi^{a}\,\chi^{b}\ge 0
\end{equation}
for any future-directed timelike vector $\chi^{a}$. For our static, spherically symmetric spacetimes, we can take $\chi^a$ to have only one non-vanishing component $\chi^t=1$. Hence, using (\ref{weak energy condition}) we obtain $T_{tt}=-a^2\,T_{t}^{t}=\rho \ge 0$. Moreover, using the energy-momentum tensor of a perfect fluid (\ref{perfect fluid energy-momentum tensor}) we find $\rho+p\ge 0$. This shows that the null energy condition (\ref{the null energy condition for the energy-momentum tensor}) is a special case of the weak energy condition. 

Adopting the weak energy condition, we can show below that $b^{\prime\prime}<0$, not only at the local extrema, but for all $r$ where $r_{\mbox{\scriptsize H}}<r<\infty$. To this end we rewrite the first equation in (\ref{tt and rr equations of motion}) in the form
\begin{equation}
\frac{d}{dr}\left(b^n\Gamma_{m,n}\right)=-b^{\prime}b^{n}T_{t}^{t}\,,
\label{compact form for the tt component}
\end{equation}
where
\begin{equation}
\Gamma_{m,n}=\frac{\alpha_{m}\,n!}{2(n-2m+1)!}b\,\left(\frac{1-a^2\,b^{\prime\;2}}{b^2}\right)^m\,.
\label{Gamma mn}
\end{equation}
 Integrating eq. (\ref{compact form for the tt component}) and using $a^2\,T_{t}^{t}=-\rho$ we obtain
\begin{equation}
\left(1-a^2\,b^{\prime\;2}\right)^m=\left(\frac{r_{\mbox{\scriptsize H}}}{b(r)}\right)^{1+n-2m}+\frac{2(n-2m+1)!}{\alpha_{m}\,n!\,b(r)^{1+n-2m}}\int_{r_{\mbox{\scriptsize H}}}^{r}d\eta\,b^n(\eta)\,b^{\prime}(\eta)\rho(\eta)/a^2(\eta)\,.
\label{the l.h.s is greather than 0}
\end{equation} 
We have shown above that the null energy condition ensures that $b(r)$ is a monotonic function and hence $b^{\prime}(r)>0$ for $r_{\mbox{\scriptsize H}}<r<\infty$, given that $b(r)>0$.  Moreover, using the weak energy condition, $\rho>0$, we see immediately that the l.h.s of eq. (\ref{the l.h.s is greather than 0}) above is always greater than zero. It is trivial to see that the same result holds for $b(r)<0$. Finally using (\ref{equation of motion of b}) proves that $b^{\prime\prime}(r)<0$ for all $r$,  $r_{\mbox{\scriptsize H}}<r<\infty$. This is analogous to the result of pure Einstein gravity eq. (\ref{equation of motion of b in Einstein gravity}) although in the latter case one uses only the null energy condition.

\section{ Entropy of pure Lovelock black holes and C-functions }
The C-functions we want to find should by definition reduce to the black hole entropy when evaluated at the horizon. We begin this section by recalling the expression for the entropy in Lovelock gravity.

A general formula for the entropy of stationary black holes in Lovelock gravity was obtained by Jacobson and Myers \cite{Jacobson:1993xs} using  Hamiltonian methods. They showed that the entropy of a black hole in pure Lovelock gravity of order $m$ is given by
\begin{equation}\label{Jacobson entropy}
S^{(m)}=4\,\pi\,m\,\alpha_{m}\int_{\mbox{\scriptsize KH}}d^{n}x\,\sqrt{-h}{\cal L} _{m-1}(h)\,,
\end{equation}
where $h$ is the induced metric on the horizon, and the integration is evaluated on any spacelike slice of the Killing horizon. Using the spherical components of the Riemann tensor $R_{ij}{}^{kl}(h)$, calculated from the induced metric on the horizon, and the basic definition of the Lagrangian in eq. (\ref{LUVLagrangian}) along with the second eq. in (\ref{second Reiman}) and the identity (\ref{recurance}), we obtain

\begin{equation}\label{final entropy}
S^{(m)}=\frac{4\,\pi\,m\,n!\,\alpha_{m}\,\Omega_{n}}{(n-2\,m+2)!}\,b^{n-2m+2}_{\mbox{\scriptsize H}}\,.
\end{equation}
Note that for $m=1$ with $\alpha_1=1/16\pi\,G_4$ we obtain $S^1=\Omega_{n}\,b^n/4\,G=A/4\,G_4$, where $A$ is the surface area of the horizon, which is the Bekenstein-Hawking expression for the entropy of Einstein gravity.

One can also use Wald's expression for the entropy (\ref{ Wald's formula}) to obtain the same result (\ref{final entropy}) above. To show this we start from the Lovelock Lagrangian $L_m$, where ${\cal L}_m=\sqrt{-g}L_{m}$ and ${\cal L}_m$ is given by (\ref{LUVLagrangian}), to obtain by direct calculations
\begin{equation}
\frac{\partial L}{\partial R_{e_1e_2}{}^{f_1f_2}}=\frac{2\,m}{2^m}\delta_{c_1d_1...c_{m-1}d_{m-1}f_1f_2}^{a_1b_1...a_{m-1}b_{m-1}e_1e_2}\,R_{a_1b_1}{}^{c_1d_1}(g)\,....\,R_{a_{m-1}b_{m-1}}{}^{c_{m-1}d_{m-1}}(g)\,.
\end{equation}
Using $\epsilon_{rt}=1$, the only nonvanishing component of the binormal to a spacial two-surface  concentric with the horizon, we find
\begin{eqnarray}
\frac{\partial L}{\partial R_{e_1e_2}{}^{f_1f_2}}\epsilon_{e_1e_2}\epsilon^{f_1f_2}=-\frac{2\,m}{2^m}\delta_{k_1l_1...k_{m-1}l_{m-1}}^{i_1j_1...i_{m-1}j_{m-1}}\,R_{i_1j_1}{}^{k_1l_1}(g)\,....\,R_{i_{m-1}j_{m-1}}{}^{k_{m-1}l_{m-1}}(g)\,,
\end{eqnarray}
where $R_{ij}{}^{kl}(g)$ denotes the spherical components of Riemann tensor calculated from the full metric. Using the last expression together with  eq. (\ref{second Reiman}) in Wald's formula we obtain
\begin{equation}\label{general entropy}
S^{(m)}=4\,\pi\,m\,\alpha_{m}\frac{n!}{(n-2\,m+2)!}\left(\frac{1-a^{2}\,b^{\prime\,2}}{b^2} \right)^{m-1}_{\mbox{\scriptsize H}}\,b^{n}_{\mbox{\scriptsize H}}\Omega_{n}\,,
\end{equation}
where  $b$ and the bracket are to be evaluated on the horizon, i.e. at $a=0$, and hence we reproduce the same result given by eq. (\ref{final entropy}).

A C-function should extend the entropy of spheres on a constant time slice away from the horizon. We see that eqs. (\ref{final entropy}) and (\ref{general entropy}) suggest different ways of doing this. This lead to two possibilities for C-functions. What we call the C-function of the first kind below is based on eq. (\ref{final entropy}), while the C-function of the second kind is based on eq. (\ref{general entropy}). The difference is that eq. (\ref{final entropy}) involves only the intrinsic  curvature of the spheres, while eq. (\ref{general entropy}) includes the extrinsic curvature as well.   
\subsection{C-functions of the first kind}

In \cite{Goldstein:2005rr} it was shown that one can take the C-function in Einstein gravity, for the static, spherically symmetric and asymptotically flat spacetimes, to be
\begin{equation}
C^{(1)}(r)=A(r)/4\,G_4=\pi\,b^2(r)/G_4\,,
\end{equation}
which coincides with the entropy at the horizon $r=r_{\mbox{\scriptsize H}}$.

\smallskip

In searching for generalizations of the C-function in higher curvature gravity, we continue to require that this function satisfies the usual properties \cite{Cremades:2006ke}, at least for the static and spherically symmetric specetimes :

\smallskip

a) It can be evaluated on any spherical surface concentric with the horizon.

b) When evaluated on the horizon of a black hole it reduces to the entropy.

c) If certain physical (e.g. null or weak) and boundary (e.g. asymptotically flat or AdS)

 $\,\,\,\,\,$ conditions are satisfied, then $C$ is a non-decreasing function along the outward 

$\,\,\,\,\,\,\,$radial direction.

\smallskip

Hence, let us take our proposed C-function of the pure Lovelock gravity of order $m$ to be equal to the expression (\ref{final entropy}) evaluated on any spherical surface concentric with the horizon, i.e. we write
\begin{equation}\label{first C function}
C^{(m)}(r)=\frac{4\,\pi\,m\,n!\,\alpha_{m}\,\Omega_{n}}{(n-2\,m+2)!}\,b^{n-2m+2}(r)\,.
\end{equation}
We have shown in the previous section that $b$ is monotonically increasing function of $r$ in pure Lovelock gravity of order $m$ as long as $\alpha_{m}>0$ and the matter content  satisfies the null energy condition. Hence $C^{(m)}$ satisfies the above three conditions and can serve as a candidate for a well defined C-function. We call these functions C-functions of the first kind
\footnote{The C-function of the first kind was noted previously without proof in \cite{Alishahiha:2006ke}.}.
 However, in the next section we will show that this is not the only C-function one can define and yet satisfy the conditions stated above.

\subsection{C-functions of the second kind}
In this section we show that another class of well defined C-functions exists. Our proposed form is motivated by expression (\ref{general entropy}) after dropping out the subscript H allowing the 
calculations of this quantity on any sphere concentric with the horizon. Hence, our second C-function takes the form
\begin{equation}\label{second C-function}
\tilde C^{(m)}(r)=\frac{4\,\pi\,m\,n!\,\alpha_{m}\,\Omega_{n}}{(n-2\,m+2)!}\left(\frac{1-a^{2}(r)\,b^{\prime\,2}(r)}{b^2(r)} \right)^{m-1}\,b^{n}(r)\,.
\end{equation}
Taking constant time slices of the metric (\ref{spherical metric}), we notice that the term $a^2\,b^{\prime\;2}$ is proportional to the extrinsic curvature of constant $r$ surfaces. Taking the normal to be $n=dr/a(r)$ we obtain
\begin{equation}
C^{(m)}\sim b^{n}\,\left(\hat R\right)^{m-1}\,,
\end{equation}
and
\begin{equation}
\tilde C^{(m)}\sim b^n\,\left(\hat R-K^2+K_{ij}K^{ij} \right)^{m-1}\,,
\end{equation}
where $\hat R$ and $K_{ij}=\nabla_{i}n_{j}$ are the intrinsic and extrinsic curvature of spheres.

Obviously, the function (\ref{second C-function}) also gives the correct form of the entropy when evaluated on the horizon, $a^2=0$. Now we turn to the question whether $\tilde C^{(m)}$ satisfies condition (c) above. In the following we show that this function, indeed, increases with radius provided that the matter content satisfies the weak energy condition.

\smallskip

 To make the notation compact, let us drop out the preceding numerical coefficients in (\ref{second C-function}) and consider instead the function
\begin{equation}
 F(r)=\kappa\,b^{n-2m+2}\left(1-a^2\,b^{\prime 2} \right)^{m-1}\,,
\end{equation} 
where $\kappa=\alpha_m\,n!/2(n-2m+1)!$.
 If this function were non-monotonic, then we could find a radius $r_0$ with $r_{\mbox{\scriptsize H}}<r_0<\infty$  such that $F^{\prime}(r_0)=0$. We start by writing the function $F$ in the form $F=b\left(b^n\,\Gamma_{m,n}\right)/\left(1-a^2\,b^{\prime\;2}\right)$, where $\Gamma_{m,n}$ is given by eq. (\ref{Gamma mn}).
By direct calculations we find that
\begin{equation}
\frac{dF}{dr}=\frac{b^{\prime}\,b^n\,\Gamma_{m,n}+b\left(b^n\,\Gamma_{m,n}\right)^{\prime}}{1-a^2\,b^{\prime\;2}}+\frac{2\,b^{n+1}\,\Gamma_{m,n}\left(a\,a^{\prime}\,b^{\prime\;2}+a^2\,b^{\prime}\,b^{\prime\prime}\right)}{\left(1-a^2\,b^{\prime\;2}\right)^2}\,.
\end{equation}
Further, we use eq. (\ref{compact form for the tt component}) and the $t-t$ component of the equations of motion (\ref{tt and rr equations of motion}) to substitute for the quantities $\left(b^n\,\Gamma_{m,n}\right)^{\prime}$ and $\left(a\,a^{\prime}\,b^{\prime\;2}+a^2\,b^{\prime}\,b^{\prime\prime}\right)$, respectively which yields
\begin{equation}
\frac{dF}{dr}=\frac{b^{\prime}\,b^n}{1-a^2\,b^{\prime\;2}}\left(\frac{n-m+1}{m}\Gamma_{m,n}+\frac{m-1}{m}b\,\rho \right)\,.
\label{ final equation to prove c function of the second kind}
\end{equation}
However, we have shown before that $b'>0$ ( b is monotonic), and the weak energy condition was enough to prove $\Gamma_{m,n}>0$. We conclude immediately that the r.h.s of (\ref{ final equation to prove c function of the second kind}) is positive definite, and hence there is no solution to $dF/dr=0$. This proves that the functions $\tilde C^{(m)}(r)$ are  C-functions for pure Lovelock gravity coupled to matter that satisfies the weak energy condition.

\smallskip

\subsection{ Example: the C-functions in the vacuum solution of pure Lovelock gravity }
We can check our results by looking at Vacuum solutions of pure Lovelock gravity. These solutions can be obtained from eq. (\ref{the l.h.s is greather than 0}) by putting $\rho=0$ , i.e. the vacuum solution is given by \cite{Crisostomo:2000bb,Cai:2006pq}
\begin{equation}
a^2(r)=1-\left(\frac{r_{\mbox{\scriptsize H}}}{r}\right)^{(n-2m+1)/m}\,\,, b(r)=r \,.
\end{equation}
The C-functions of the first and second kind for these spacetimes are given by
\begin{eqnarray}
\nonumber
C^{(m)}&=&\frac{4\,\pi\,m\,n!\,\alpha_{m}\Omega_{n}}{(n-2m+2)!}r^{n-2m+2}\,,\\
\nonumber
\tilde C^{(m)}&=&C^{(m)}\left(\frac{r_{\mbox{\scriptsize H}}}{r}\right)^{(n-2m+2)(m-1)/m}\,,\\
 &=&\frac{4\,\pi\,m\,n!\,\alpha_{m}\Omega_{n}}{(n-2m+2)!}r^{(n-2m+2)/m}r_{\mbox{\scriptsize H}}^{(n-2m+2)(m-1)/m}\,.
\end{eqnarray}
 From the above equations we see that both $C^{(m)}$ and $\tilde C^{(m)}$ are monotonic functions of the radial coordinate, and both reduce to the entropy when evaluated on the horizon. We also see that $\tilde C^{(m)}<C^{(m)}$ for all $r>r_{\mbox{\scriptsize H}}$.

\section{ General Lovelock gravity, entropy and C-functions }
Our results in section 3 were for pure Lovelock theories, with the coefficient of only one of the Lovelock terms in the Lagrangian nonzero. Now we want to ask whether these results hold in a general Lovelock gravity theory.

The Lagrangian for general Lovelock gravity is given by ${\cal L}=\sum_{m=1}^{[D/2]}\alpha_{m}\,{\cal L}_{m}$, where ${\cal L}_{m}$ are given by eq. (\ref{LUVLagrangian}), and we drop the cosmological constant as we are interested in asymptotically flat solutions. The equations of motion read 
\begin{equation}\label{general equations of LUV}
{\cal G}_{e}{}^{f}=\sum_{m=1}^{[(D-1)/2]}\alpha_{m}\,{\cal G}_{e}{}^{f\,(m)}/2^{m+1}=-T_{e}{}^{f}\,.
\end{equation}
Similarly the entropy  is given by $S=\sum_{m=1}^{[(D-1)/2]}S^{(m)}$, with $S^{(m)}$  given by  eq. (\ref{final entropy}).

In the previous section we proved that $b$ is a monotonic function of $r$ in all pure Lovelock theories. We can show that this result still holds for general Lovelock theories. As before, we take the combination
\begin{equation}
\sum_{m=1}^{[(D-1)/2]}\alpha_{m}\left({\cal G}_t{}^{t\,(m)}-{\cal G}_r{}^{r\,(m)}\right)/2^{m+1}=T_{r}^{r}-T_{t}^{t}=T_{ab}\xi^{a}\,\xi^{b}\,,
\end{equation}
and using eq. (\ref{tt and rr equations of motion}) we obtain
\begin{equation}
a^2\,b^{\prime\prime}/b^2=-\frac{T_{ab}\xi^{a}\xi^{b}}{ \sum_{m=1}^{[(D-1)/2]}  \frac{mn!\,\alpha_{m}}{(n-2m+1)!}\left(\frac{1-a^2\,b^{\prime\,2}}{b^2} \right)^{m-1}    }\,.
\end{equation}
  Assuming the positivity of the coupling constants $\alpha_{m}$ for all orders, and using the same reasoning as in pure Lovelock gravities, we conclude that $b$ is also monotonic in general Lovelock gravity coupled to matter that satisfies the null energy condition.

\subsection{ C-functions in general Lovelock gravity}

For general Lovelock gravity, one can define either the general C-function of the first or second kind. In the first case the C-function is given by
\begin{equation}
C=\sum_{m=1}^{[(D-1)/2]}C^{(m)}=4\,\pi\,n!\,\Omega_{n}\,b^{n+2}(r)\,\sum_{m=1}^{[(D-1)/2]}\frac{m\,\alpha_{m}\,b^{-2m}(r)}{(n-2m+2)!}\,.
\end{equation} 
Taking the derivative w.r.t $r$ we obtain
\begin{equation} 
C^{\prime}\propto b^{\prime}\,b^{n+1}\sum_{m=1} \frac{m\alpha_{m}\,b^{-2m}}{(n-2m+1)!}\,.
\end{equation}
  As we showed before $b^{\prime} \ne 0$ for $r_{\mbox{\scriptsize H}}<r<\infty$. Moreover, the positivity of the coupling constants $\alpha_{m}$ ensures that there is no solution to the polynomial under the sum. This proves the monotonicity of the general C-functions of the first kind provided that the matter content satisfies the null energy condition.

\smallskip

In the same way we define the general C-function of the second kind to be 
\begin{equation}
\tilde C=\sum_{m=1}^{[(D-1)/2]} \tilde C^{\,(m)}\,.
\end{equation}
 However, testing the monotonicity of this function is generally complicated. In the following we restrict our analysis to the case of Gauss-Bonnet gravity.

\subsection{ C-function of the second kind in Gauss-Bonnet gravity}

The C-function of the second kind in $D=n+2$ dimensional Gauss-Bonnet gravity reads 
\begin{equation}\label{second class c function in gauss-bonnet gravity}
\tilde C_{\mbox{\scriptsize GB}}=4\,\pi\,\Omega_{n}\left[\alpha_{1}\,b^n+2\,n\,(n-1)\,\alpha_{2}\,b^{n-2}\,\left(1-a^2\,b^{\prime\,2}\right) \right]\,.
\end{equation}
 To prove the monotonicity of this function we proceed as we did before. We define the function
\begin{equation}
F(r)=b^n+2\,n\,(n-1)\,\alpha\,b^{n-2}\,\left(1-a^2\,b^{\prime\,2}\right)\,,
\end{equation}
 where $\alpha=\alpha_{2}/\alpha_{1}$. We then ask whether it is possible to find solutions to $dF/dr=0$ where
\begin{equation}\label{ derivative for f in GB}
\frac{dF}{dr}=n\,b^{\prime}\,b^{n-3}\left[b^2+2\,(n-1)\alpha\left((n-2)\left(1-a^2\,b^{\prime\,2} \right)-2b\left(a^{2}\,b^{\prime\,\prime}+a\,a^{\prime}\,b^{\prime}\right) \right) \right]\,.
\end{equation}
Using the equations of motion (\ref{tt and rr equations of motion}) and (\ref{general equations of LUV}), the relation $a^2\,T_{rr}+T_{tt}/a^2=T^{ab}\xi_{a}\xi_{b}$, with $\xi_{a}$ being a null vector, and the definition of $F$ above we find
\begin{equation}
a^2\,b^{\prime\prime}+a\,a^{\prime}\,b^{\prime}=\frac{\left[b(F-b^n)/4+(n-2)(n-3)(F-b^n)^2/8n(n-1)b^{n-1}-n!\,\alpha\,b^{n+1}\,T_{tt}/a^2 \right]}{n\,\alpha\,b^n+(n-2)\,\alpha(F-b^n)}\,.
\end{equation}
Substituting this expression into eq.(\ref{ derivative for f in GB}), we find the the solutions of $F^{\prime}=0$ are given by the solutions of the equation
\begin{equation}
F^2-\frac{6\,b^n}{(n-1)(n-2)}F+\left[\frac{(n+1)(n+2)}{(n-1)(n-2)}+\frac{8\,n\,n!\,\alpha}{(n-2)}\frac{T_{tt}}{a^2} \right]b^{2n}=0\,,
\end{equation} 
from which we see immediately that there are no real solutions for $n \ge 3$ if $\alpha>0$, and  $T_{tt}=\rho \ge 0$, i.e. for matter content that satisfies the weak energy condition. This proves the monotonicity of the C-function of the second kind in Gauss-Bonnet gravity. 

\smallskip

Going to higher order Lovelock gravity, at least according to the present method, requires thorough analysis of higher degree polynomials. We will not attempt to carry this out here. Instead, in the next section we use numerical techniques to study the monotonicity of the C-function of the second kind. For a particular type of matter satisfying the weak energy condition, we will verify that $\tilde C$ is monotonic for Gauss-Bonnet and see that it is also monotonic including the third order Lovelock term with positive coefficient. This result suggests that it might be possible to improve on the results in this section and show monotonicity of $\tilde C(r)$ for all Lovelock gravity theories with coefficients $\alpha_{m} >0$.

\subsection{Numerical example: Gauss-Bonnet and third order Lovelock gravity}

In this section we work out various numerical examples that show the monotonicity of C-functions of the second  kind. 
In the following we consider general Lovelock gravity in $D$ dimensions coupled to two abelian gauge fields $A_{a}^{\mu}$ with $a=1,2$ and a massless scalar modulus field $\phi$. This has been recentely studied in the context of the attractor mechanism \cite{Goldstein:2005hq,Anber:2007gk}.
 The modulus scalar has vanishing potential, but couples to the gauge field kinetic terms through the matrix function $f_{ab}(\phi)$. The action is given by
\begin{equation}
S=\int dx^D\,\sqrt{-g}\left[\sum_{m=1}^{[D/2]}\alpha_{m}\,{\cal L}_{m}-2\,\partial_{\mu}\phi\partial^{\mu}\phi-f_{ab}(\phi)F_{\mu\nu}^{a}F^{b\,\mu\nu} \right]\,,
\end{equation}
where $\mu\,,\nu=0,...,D-1$. Using the static and spherically symmetric ansatz (\ref{spherical metric}), we look for solutions to the resulting equations of motion (\ref{general equations of LUV}), where their explicit form is given for 5-$D$ in a previous work \cite{Anber:2007gk}. The equations of motion for the gauge fields $\partial_{\mu}\left(\sqrt{-g}\,f_{ab}F^{a \,\mu\,\nu}\right)=0$ may be solved by taking the field strengths to be of the form
\begin{equation}
F^{a}=\frac{f^{ab}\,Q_{b}}{b^n}dt\wedge dr\,,
\end{equation}   
where $Q_b$ are the electric charges, and the field dependent tensor $f^{ab}(\phi)$ is the inverse of the tensor coupling $f_{ab}(\phi)$ that appears in the Lagrangian. With this form for the field strength, the energy-momentum tensor for the gauge fields can be written in terms of the effective potential
\begin{equation}
V_{\mbox{\scriptsize eff}}=f^{cd}(\phi)\,Q_{c}\, Q_{d}\,,
\end{equation}
and hence
\begin{eqnarray}
\nonumber
T_{t}^t&=&-a^2\,(\partial_r \phi)^2-\frac{V_{\mbox{\scriptsize eff}}}{b^{2n}}\,,\\
T_{r}^r&=&a^2\,(\partial_r \phi)^2-\frac{V_{\mbox{\scriptsize eff}}}{b^{2n}}\,.
\end{eqnarray}
Also, the effective potential acts as a potential in the equation of motion for the modulus scalar, which is given by
\begin{equation}\label{equation of motion of the modulus field}
\partial_{r}\left(b^n\,a^2\,\partial_{r}\phi\right)=\frac{V^{\prime}_{\mbox{\scriptsize eff}}(\phi)}{2\,b^n}\,.
\end{equation}  
This equation may be solved by a constant value $\bar{\phi}$ of the scalar field if this value represents a critical point of the effective potential, i.e. $V^{\prime}_{eff}(\bar\phi)=0$. Given that the scalar field  is constant throughout the spacetime one obtains a simple solution to the equations of motion, namely summing over all Lovelock orders $m$ in eq. (\ref{compact form for the tt component}) one obtains non-extremal black hole solution whose metric functions are solutions of the equation                      
\begin{equation}\label{non exteremal BH solution in GB }
\sum_{m=1}^{[(D-1)/2]}\frac{n!\,\alpha_{m}\,r_{\mbox{\scriptsize H}}^{2n-2m}}{2(n-2m+1)!}\left[(1-a^2(r))^m\left(\frac{r}{r_{\mbox{\scriptsize H}}}\right)^{1+n-2m}-1 \right]=\frac{V_{\mbox{\scriptsize eff}}(\bar\phi)}{n-1}\left[1-\left(\frac{r}{r_{\mbox{\scriptsize H}}}\right)^{n-1}\right]\,,
\end{equation}
and $b(r)=r$, where $r_{\mbox{\scriptsize H}}$ is the outer horizon radius of the black hole. 
\footnote{For explicit expression in Gauss-Bonnet gravity see \cite{Anber:2007gk}. }

\smallskip 

Now we want to consider solutions with non-constant $\phi$. To this end, we take small perturbations of the scalar field near the horizon where the metric functions are approximated by 
\begin{equation}
a^2(r)\approx\rho(r-r_{\mbox{\scriptsize H}})\,,\,\,\,\,\, b(r)\approx r_{\mbox{\scriptsize H}}\,,
\end{equation}
and $\rho$ is given by
\begin{equation}
\rho=\left(\frac{\partial a^2}{\partial r}\right)_{r_{\mbox{\scriptsize H}}}=\frac{S_1-V_{\mbox{\scriptsize eff}}(\bar \phi)/r_{\mbox{\scriptsize H}}^n}{ S_2   }   \,,
\end{equation}
where
\begin{eqnarray}
\nonumber
S_1&=&\sum_{m=1}^{[(D-1)/2]}\frac{n!\alpha_{m}r_{\mbox{\scriptsize H}}^{n-2m}}{2(n-2m)!}\,,\\
S_2&=&\sum_{m=1}^{[(D-1)/2]}\frac{n!\,m\alpha_{m}r_{\mbox{\scriptsize H}}^{1+n-2m}}{2(n-2m+1)!}\,.
\end{eqnarray}

Considering the scalar field perturbation $\phi(r)=\bar\phi+\epsilon\,\phi_{1}(r)$, where $\epsilon<<1$, we find that the first order perturbative equation in the near horizon region is then given by
\begin{equation}
(r-r_{\mbox{\scriptsize H}})\phi_{1}^{\prime\prime}+\phi_{1}^{\prime}-\frac{\beta^2}{2\,r_{H}^{2n}\rho}\phi_{1}=0\,,
\end{equation}
where $\beta^2=V^{\prime\prime}(\bar\phi)$. The well behaved solution for linearized perturbations of the scalar field is then given by
\begin{equation}\label{solution for phi1}
\phi_{1}(r)=E\,I_{0}\left[\frac{\beta}{r_{\mbox{\scriptsize H}}^n}\sqrt{\frac{2(r-r_{\mbox{\scriptsize H}})}{\rho}} \right]\,,
\end{equation}
where $I_0$ is the modified Bessel's function of the first kind and $E$ is an integration constant.

\smallskip

 To this end, one can use the solution to the scalar field perturbation as initial condition to the full non-linear system. In the following we numerically integrate the non-linear equations (\ref{general equations of LUV}) and (\ref{equation of motion of the modulus field})  using the Rung-Kutta method.  We also take the couplings of the scalar field to the gauge fields to be 
\begin{equation}
f_{ab}(\phi)=e^{-\gamma_{a}\phi}\delta_{ab\,,}
\end{equation}  
from which we find immediately that the effective potential is given by
\begin{equation}
V_{\mbox{\scriptsize eff}}(\phi)=e^{\gamma_{1}\,\phi}Q_{1}^2+e^{\gamma_{2}\,\phi}\,Q_2^2\,.
\end{equation}

\begin{figure}[ht]
\leftline{
\includegraphics[width=0.5\textwidth]{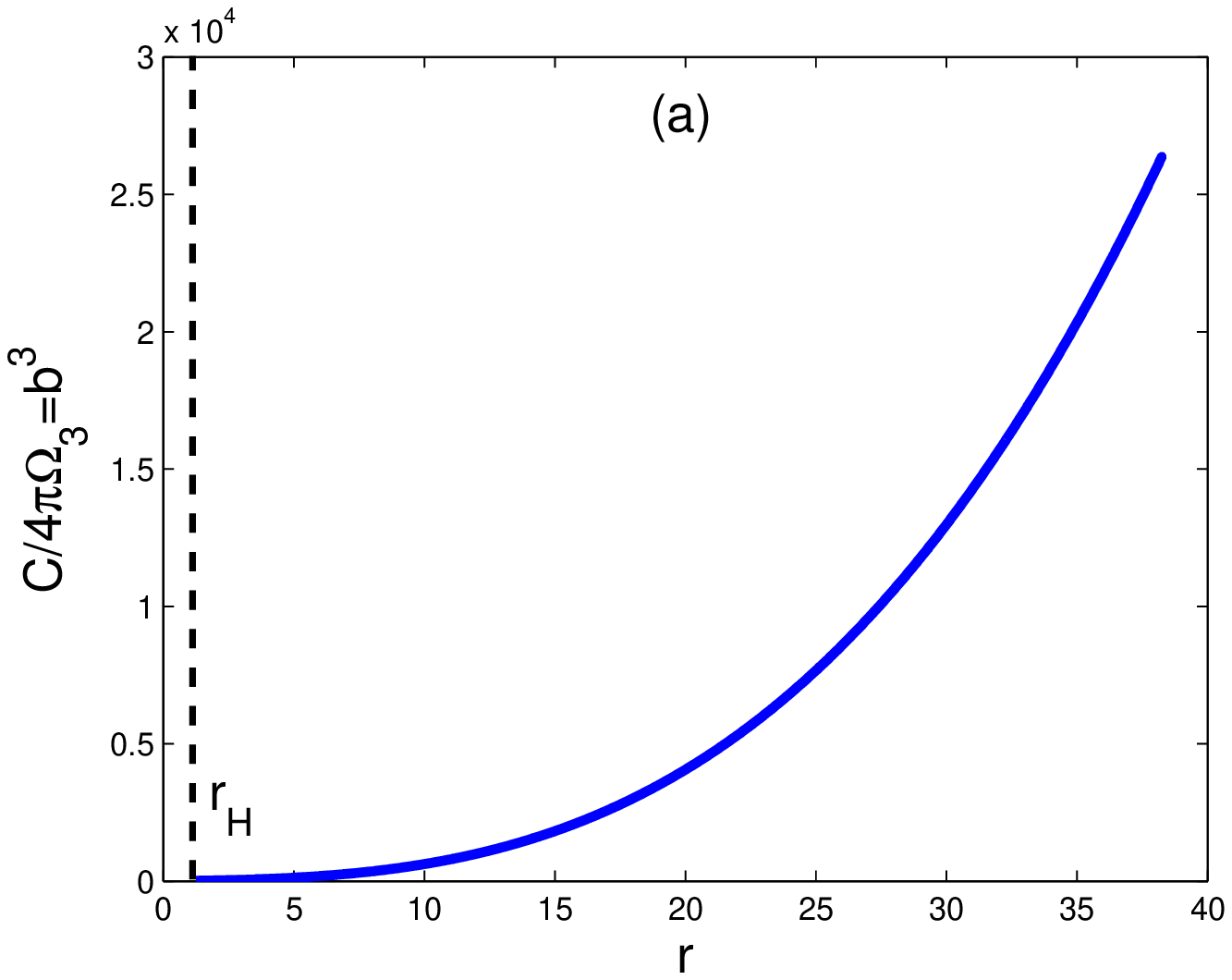}
\includegraphics[width=0.5\textwidth]{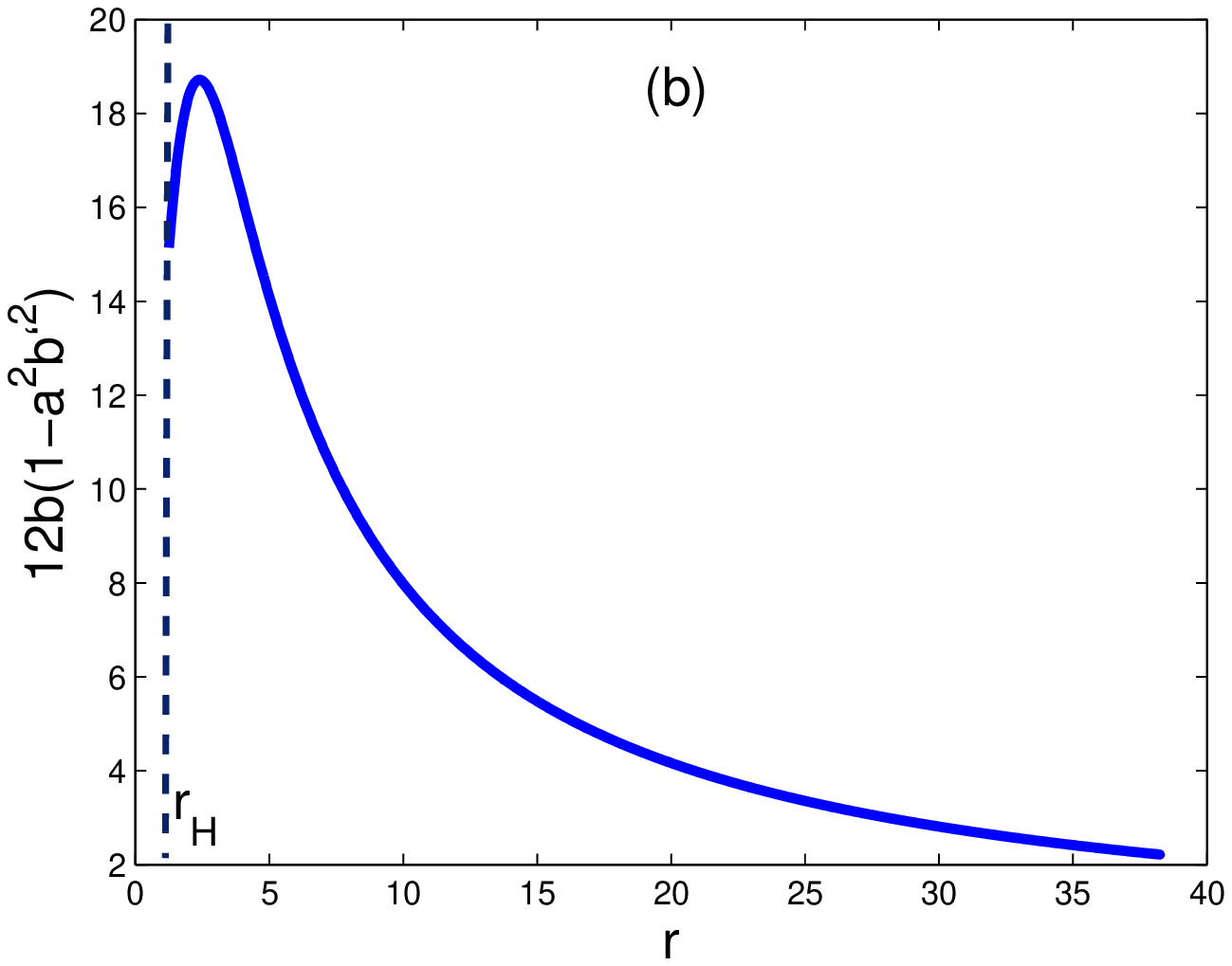}
}
\centerline{
\includegraphics[width=0.5\textwidth]{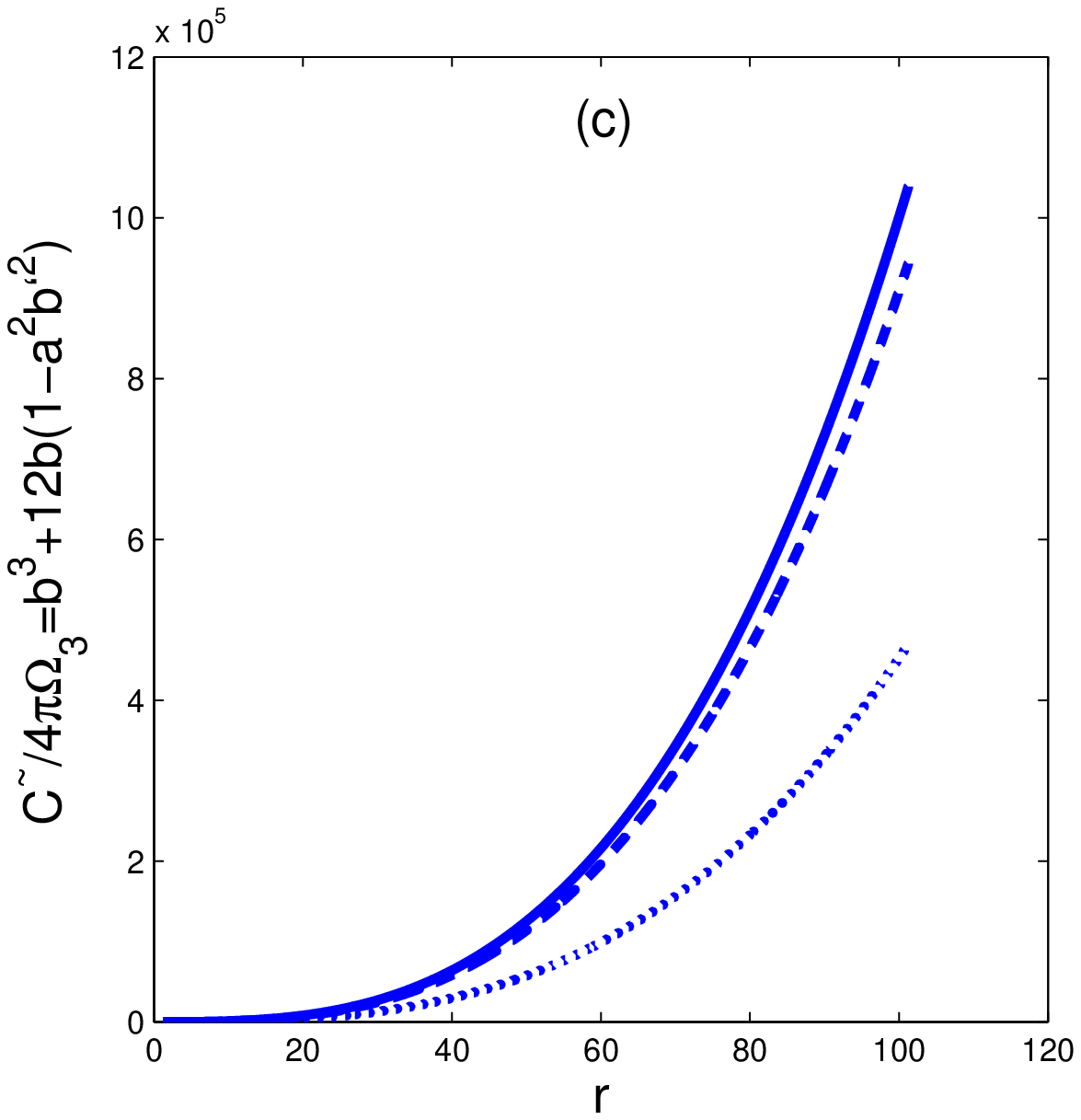}
}

\caption{
 Numerical results for the C-functions of the first and second kind in $D=5$ Gauss-Bonnet gravity. We choose $r_{\mbox{\scriptsize H}}=1.26$, coupling constants $\alpha_1=\alpha_2=1$, charges $Q_{e1}=1/\sqrt{2}$ and $Q_{e2}=\sqrt{2}$, $\gamma_1=-\gamma_2=2.0$ and  $\delta r=0.01$ in our numerical scheme. We also take $E=0.4$ in the first two figures, while in the third figure we use  $E=0.0\,,0.2\,,0.4$ for the solid, dashed and dotted lines respectively.
}
\label{compare C in GB}
\end{figure}

\begin{figure}[ht]
\leftline{
\includegraphics[width=0.5\textwidth]{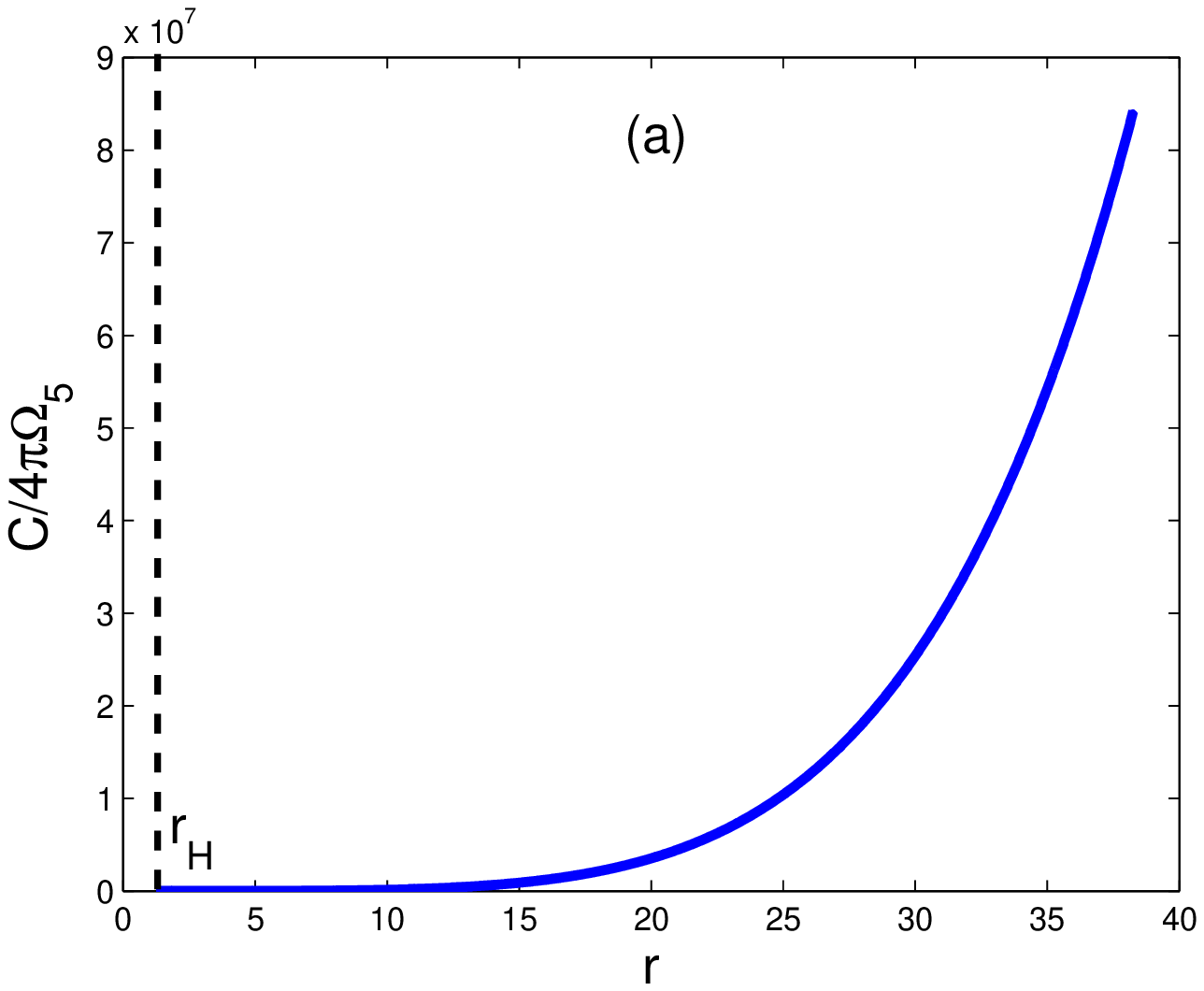}
\includegraphics[width=0.5\textwidth]{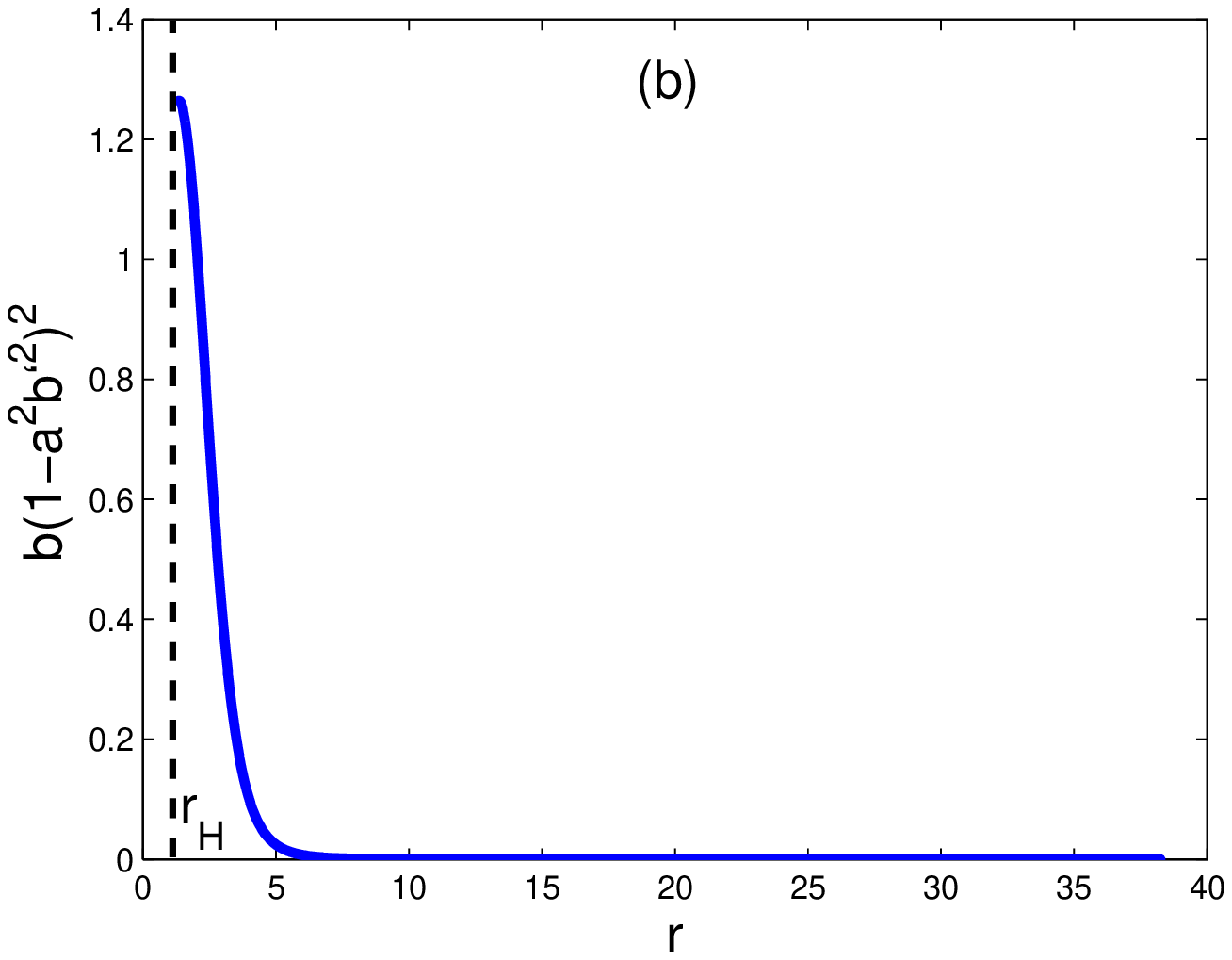}
}
\centerline{
\includegraphics[width=0.5\textwidth]{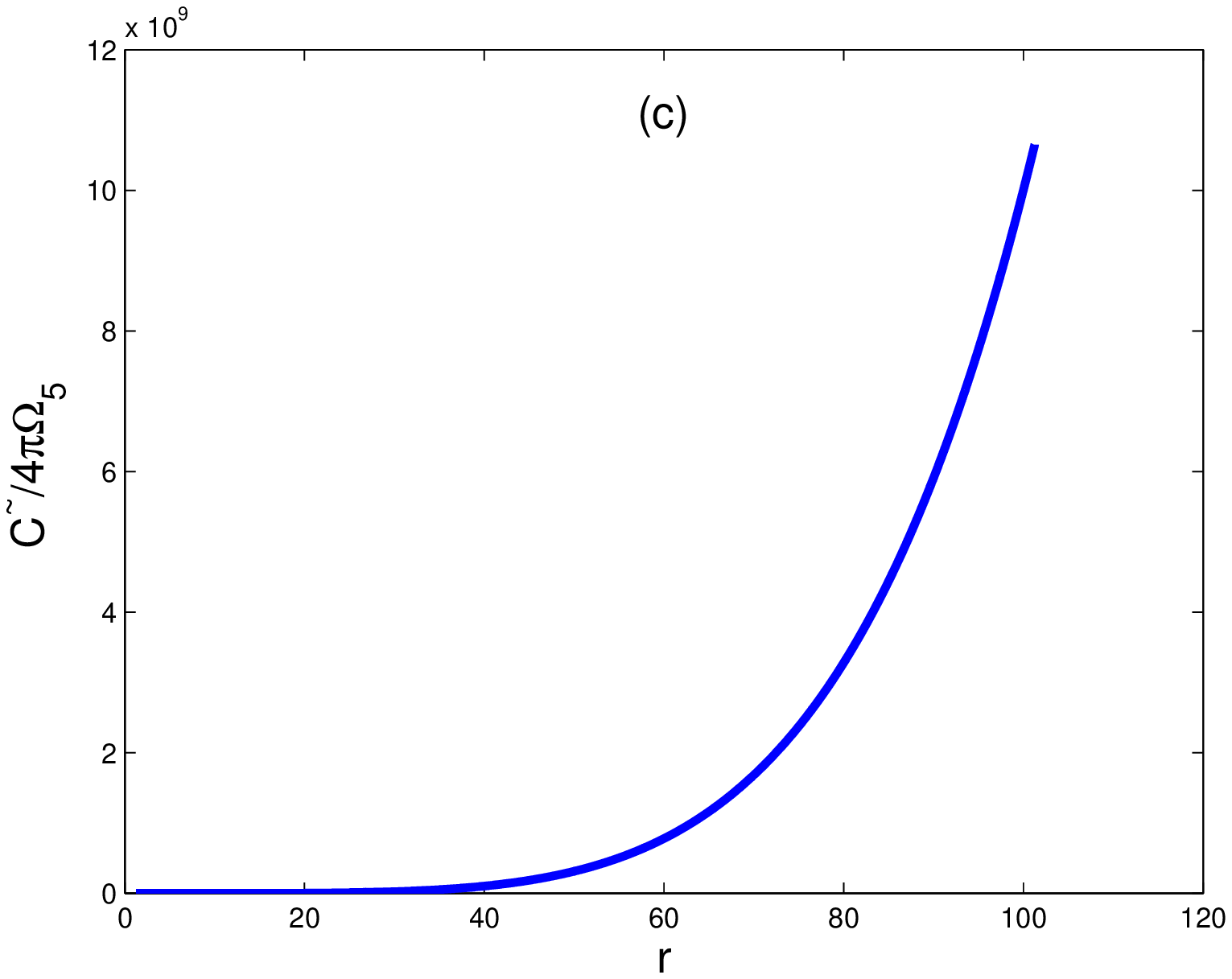}
}

\caption{
 Numerical results for the C-functions of the first and second kind in $D=7$ third order Lovelock gravity. We choose $r_{\mbox{\scriptsize H}}=1.26$, coupling constants $\alpha_1=\alpha_2=\alpha_3=1$, charges $Q_{e1}=1/\sqrt{2}$ and $Q_{e2}=\sqrt{2}$, $\gamma_1=-\gamma_2=2.0$. We also take $\delta r=0.01$ and $E=0.4$  in our numerical scheme.}
\label{compare C in 3rd LUV}
\end{figure}

In figure (\ref{compare C in GB}) we compare the C functions of the first and second kind in Gauss-Bonnet gravity in five dimensions. We choose $r_{\mbox{\scriptsize H}}=1.26$, coupling constants $\alpha_1=\alpha_2=1$, charges $Q_{e1}=1/\sqrt{2}$ and $Q_{e2}=\sqrt{2}$, $\gamma_1=-\gamma_2=2.0$. We denote the proximity to the horizon by the parameter $\delta r=(r_i-r_{\mbox{\scriptsize H}})$/$r_i$, this is where the initial conditions are set using the pertarbative results above, and we take it to be $0.01$ in our numerical scheme. We also take $E=0.4$ in the first two figures.
 Figure (a) shows the monotonic behavior of the  C-functions of the first kind. On the contrary, Figure (b) shows the non-monotonic behavior of the term $12\,b(1-a^2\,b^{\prime\,2})$ that appears in the C-function of the second kind  in eq. (\ref{second class c function in gauss-bonnet gravity}). However, as it is clear form figure (c), when we add up the $b^3$ term the overall function restores its monotonic behavior. In addition, In figure (c) we compare the C-function for different values of the constant $E$, we take $E=0.0\,,0.2\,,0.4$ for the solid, dashed and dotted lines respectively. We notice that the ultraviolet value of $\tilde C$ decreases as we increase the value of the constant $E$, or in other words as we increase the asymptotic value of the scalar modulus. 

In figure (\ref{compare C in 3rd LUV}) we show the results for the case of third order Lovelock gravity in seven dimensional spacetime. We see a similar behavior to the case of $D=5$: although the third term in  $\tilde C$ is decreasing, the overall function is monotonically increasing in $r$.

 Higher order Lovelock terms have also been considered numerically up to the fifth order, and all results show monotonic behavior for $\tilde C (r)$.

\section{Conclusion}

In this paper we have constructed two different C-functions for the static, spherically symmetric black holes in Lovelock gravity. This construction was inspired by Wald's expression for the entropy of stationary black holes applied to  Lovelock gravity. Although this expression is given as an integral over the induced metric on the Killing horizon, we were able to show that extending this expression in two different ways, by evaluating it on any spherical surface concentric with the horizon, gives the desired C-functions. These functions have different asymptotic values, but they degenerate at the horizon to the entropy of the black hole.

 In the case of pure Lovelock gravity of order $m$, the expression of the C-function of the first kind is simply proportional to $b^{n-2m+2}$, while that of the second kind is given by the former expression multiplied by the factor $(1-a^2\,b^{\prime\,2})^{m-1}$ which was shown to contain contributions from the  extrinsic curvatures of an $n$-sphere embedded in $n+1$ dimensional space. We have also proven the monotonicity of these functions provided that the sapce is asymptotically flat, and the matter content satisfies the null and the weak energy conditions for the first and second C-functions, respectively. It is worth mentioning that one can still show the monotonicity of the C-function of the first kind if we replace the asymptotically flat by asymptotically AdS space since the later satisfies the null energy condition.

In a general Lovelock gravity, It is natural to expect that the C-functions can be obtained by summing over pure C-functions of different orders. Although we proved the monotonicity of the general C-function of the first kind, we proved the monotonicity of the C-function of the second kind only in the case of Gauss-Bonnet gravity. It is still not obvious how to give a similar proof in the case of higher order gravity. However, we have given a numerical example in third order Lovelock gravity that indicates that the monotonicity of $\tilde C (r)$ may still hold in general. We have also checked numerically that the results hold in similar examples in higher order Lovelock theories. It is worth noting that a quasi-local mass function in Gauss-Bonnet gravity was defined in \cite{Maeda:2007uu} which also exhibits the monotonicity behavior under the dominant energy condition.

The existence of two different C-functions raises the question whether there is some reason to prefer one over the other on physical grounds. The answer to this question relies on the existence of a covariant formulation of the C-function, which may reduce to one of the C-functions defined in this work, when evaluated in static and spherically symmetric spacetimes. In turn, the existence of such a covariant function would establish a second law of black hole mechanics (which was established for Einstein gravity in \cite{Hawking:1971vc} ) in Lovelock gravity: if certain energy condition is satisfied, then the entropy of a black hole can never decrease. In other words, it may be that the second law of black hole mechanics selects the C-function that respects the law. Work along these lines is in progress.

\subsection*{Acknowledgment}

This work was supported in part by NSF grant PHY-0555304.

\end{document}